\documentclass[amsmath,amssymb,pra,twocolumn]{revtex4}
\usepackage[cp1251]{inputenc}
\usepackage[english]{babel}
\usepackage{color}
\usepackage{graphicx}
\usepackage{dcolumn}
\usepackage{braket}
\usepackage{amsmath}

\usepackage{natbib}
\begin{document}

\bibliographystyle{unsrt}

\title{Azimuthal entanglement and multichannel Schmidt-type decomposition of non-collinear biphotons}

\author{M.V. Fedorov}
\email{fedorovmv@gmail.com}
\address{A.M.~Prokhorov General Physics Institute,
 Russian Academy of Sciences, 38 Vavilov st., Moscow, 119991, Russia}

\date{\today}

\begin{abstract}
Purely azimuthal entanglement is analyzed for noncollinear frequency-degenerate biphoton states. The degree of azimuthal entanglement is found to be very high, with the Schmidt parameter $K$ on the order of the ratio of the pump waist to its wavelength. A scheme is suggested for partial realization of this high entanglement resource in the form of a multichannel Schmidt-type decomposition.
\end{abstract}

\maketitle
\bibliographystyle{unsrt}

\section{Introduction}

A structure of emission in the type-I Spontaneous Parametric Down-Conversion (SPDC) is well known: SPDC photons propagate along a cone with the axis ($0z$) coinciding with the central propagation direction of the pump, and section of the cone by the plane $(xy)\perp 0z$ is a ring \cite{Kwiat-94,Shih,Kim}. As quantum objects, SPDC photons are characterized by their wave function depending on transverse components of wave vectors ${\vec k}_{1\,\perp}$ and ${\vec k}_{2\,\perp}$, where the indices 1 and 2 indicate two indistinguishable SPDC photons and $\perp$ refers to the plane $(xy)$. Each of two SPDC photons has two degrees of freedom, for example, corresponding to motions in $0x$ and $0y$ directions. In this specific case the biphoton wave function depends on two pairs of variables, $k_{1,2\,x}$ and $k_{1,2\,y}$. Alternatively, the wave vectors ${\vec k}_{1,2\,\perp}$ can be characterized by their absolute values  $\rho_{1,2}=|{\vec k}_{1,2\,\perp}|$ and angles with respect to the $x$-axis (azimuthal angles), $\alpha_{1,2}$. This parametrization of transverse  wave vectors ${\vec k}_{1,2\,\perp}$ is widely used in the analysis based on the concept of the Orbital Angular Momentum of photons (OAM) \cite{Torres, Pudgett-10, Miatto, Giovannini}. Parametrization used in the present work slightly differs from that used in the OAM analysis. The transverse wave vectors ${\vec k}_{1,2\,\perp}$ are assumed to be characterized by spherical angles of the total wave vectors ${\vec k}_i$, i.e., by polar (zenith) angles $\theta_{1,2}$ defined as angles between ${\vec k}_i$  and the $z$-axis, and azimuthal angles $\alpha_{1,2}$ defined as previously as angles between ${\vec k}_{i\,\perp}$ and the $x$-axis. In terms of these definitions one can investigate separately entanglement of noncollinear boiphotons either in polar or in azimuthal angular variables. As far as I know, this formulation of the problem differs from that used in the OAM analysis where azimuthal and ``radial" entanglements (in variables $\alpha_{1,2}$ and $\rho_{1,2}$) are considered usually as inseparable parts of the total biphoton entanglement. As argued below, consideration of azimuthal entanglement itself has sense both for theoretical analysis and, potentially, for practical experimental researches. For the cases of sufficiently pronounced degree of noncollinearity, the degree of azimuthal entanglement will be shown to be very high and, roughly, determined by a large parameter of the pump waist divided by its wavelength. In principle, this provides a very large resource of azimuthal entanglement, and a scheme for its partial realization in experiments will be described.

\section{Biphoton angular wave function}

\subsection{General expressions}

 Let us consider a Biphoton State (BS) formed in the noncollinear frequency-degenerate process of Spontaneous Parametric Down-Conversion (SPDC) with the phase matching of the type-I. In this case the pump is propagating in a nonlinear crystal as the extraordinary wave, and some photons of the pump decay for two indistinguishable photons ``1" and ``2" of the ordinary wave, $e\rightarrow o+o$. The emitted photons are assumed to have coinciding frequencies equal to the half of a given frequency of the pump, $\omega_1=\omega_2=\omega_p/2$, and coinciding (e.g., horizontal) polarizations. Let the pump be propagating along the $z$-axis and having the waist $w$. As usually assumed, let the pump waist $w$ be much smaller than the transverse sizes of a crystal, which provides the transverse-momentum conservation:
\begin{equation}
 \label{transv-conserv}
 {\vec k}_{p\,\perp}={\vec k}_{1\,\perp}+{\vec k}_{2\,\perp},
\end{equation}
where ${\vec k}_{p\,\perp}$ is the projection of a pump wave vector ${\vec k}_p$ on the plane $(xy)$ perpendicular to the $z$-axis.
In the transverse-momentum representation the wave function of two emitted photons is known to have the form \cite{Monk,Law-E,2007,2008}
\begin{equation}
  \Psi\propto\exp\left[-\frac{w^2}{2}\left({\vec k}_{1\perp}+{\vec k}_{2\perp}\right)^2 \right]{\rm sinc}\left(\frac{L\Delta}{2}\right),
 \label{Psi}
\end{equation}
where the pump amplitude is taken Gaussian, ${\rm sinc}(x)=\sin x/x$, $L$ is the length of a crystal in the pump-propagation direction, and the phase mismatch $\Delta$ is given by
\begin{equation}
 \label{Delta}
 \Delta=k_{p\,z}-k_{1\,z}-k_{2\,z}\approx k_p-k_1-k_2+ \frac{\left({\vec k}_{1\perp}-{\vec k}_{2\perp}\right)^2}{4k_1}.
\end{equation}

\subsection{Spherical angles}
\begin{figure}[h]
\centering\includegraphics[width=6 cm]{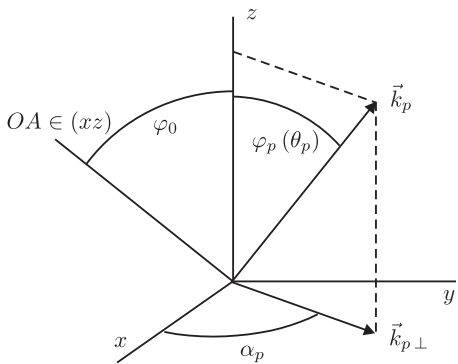}
\caption{{\protect\footnotesize {Azimuthal ($\alpha_p$) and polar ($\varphi_p(\theta_p)$) angles of the wave vector ${\vec k}_p$ ($\varphi_p$ in and $\theta_p$ outside of the crystal); $\varphi_0$ is the angle between the crystal's major optical axis and the central propagation direction of the pump (the $z$-axis).}}}\label{Fig1}
\end{figure}

In accordance with the goals declared in the Introduction, let us characterize orientation of wave vectors in a free space after the crystal by their spherical angles - polar angles $\theta_{p,1,2}$ and azimuthal angles $\alpha_{p,1,2}$:
\begin{equation}
 \label{kp-ki-sph-coord}
 \begin{matrix}
  {\vec k}_p=\frac{2\pi}{\lambda_p}\{\sin\theta_p\cos\alpha_p,\,\sin\theta_p\sin\alpha_p,\,\cos\theta_p\},\\
  {\vec k}_1=\frac{\pi}{\lambda_p}\{\sin\theta_1\cos\alpha_1,\,\sin\theta_1\sin\alpha_1,\,\cos\theta_1\},\\
  {\vec k}_2=\frac{\pi}{\lambda_p}\{-\sin\theta_2\cos\alpha_2,\,-\sin\theta_2\sin\alpha_2,\,\cos\theta_2\},
 \end{matrix}
\end{equation}
where $\lambda_p$ is the pump wavelength.
 As an example, the polar and azimuthal angles of the pump wave vector are sown in Fig. \ref{Fig1},
with the coordinate frame turned intentionally for clearer visibility in such a way that $z$-axis is vertical, though traditionally and in all further illustrations the propagation axis $0z$ is taken horizontal. Two deferent notations, $\varphi_p$ and $\theta_p$, are used for polar angles of ${\vec k}_p$ inside and outside of a crystal. The similar pairs of notations are used below for polar angles of SPDC photons ${\vec k}_{1,2}$: $\varphi_{1,2}$ and $\theta_{1,2}$.

\begin{figure}[h]
\centering\includegraphics[width=6 cm]{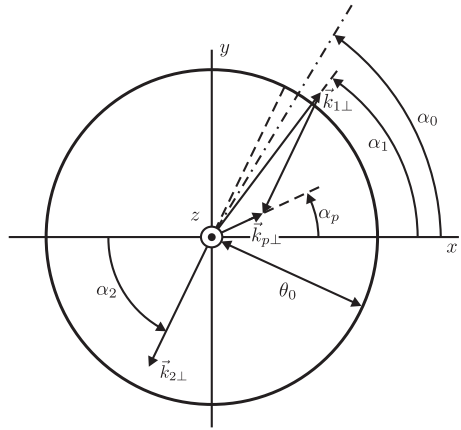}
\caption{{\protect\footnotesize {Azimuthal angles of emitted photons $\alpha_1$ and $\alpha_2$, and the azimuthal angle $\alpha_p$ of the pump wave vector; $\alpha_0=\frac{1}{2}(\alpha_1+\alpha_2)$; $\theta_0$ is the opening angle of the emission cone; the ring is the section of the cone by the plane $(xy)\perp 0z$.}}}\label{Fig2}
\end{figure}

The signs $``-"$ in the definition of ${\vec k}_2$ in Eq. (\ref{kp-ki-sph-coord}) deserve a special explanation. It's known \cite{Kwiat-94,Shih,Kim} that in the noncillinear SPDC procees two photons of each pair are located approximately at the opposite ends of diameters of the ring formed by section of the emission cone by the transverse plane $(xy)\perp 0z$. This last condition is provided in different ways in the 2D and $3D$ geometries. In the plane geometry deviations from the $z$-axis can be characterized by angles $\theta_{1,2}$ having different signs for ``up$"$ and ``down$"$ (or ``left$"$ and ``right$"$) deviations \cite{Scr}. In contrast, in the $3D$ geometry, both polar angles $\theta_1$ and $\theta_2$ are always positive ($\pi\geq\theta_{1,2}\geq 0$). In this case the condition of location at the opposite ends of the ring diameters is provided by different definitions of the azimuthal angles of two photons: these angles have to differ from each other approximately by the term $\pi$. But in the definitions used in Eq. (\ref{kp-ki-sph-coord}) and below this difference is already taken into account: if the angle $\alpha_1$ is defined as counted from the positive direction of the $x$-axis, the angle $\alpha_2$ is counted from its negative direction (see Fig. \ref{Fig2}). This shift for $\pi$ gives rise to the sings $``-"$ in the definition of ${\vec k}_2$ in Eq. (\ref{kp-ki-sph-coord}).

The representation of wave vectors in terms of spherical angles (\ref{kp-ki-sph-coord}) can be used to find rather simple expressions for the squared difference and sum of ${\vec k}_{1\,\perp}$ and ${\vec k}_{2\,\perp}$ in Eq. (\ref{Delta}):
\begin{gather}
 \nonumber
 |{\vec k}_{1\,\perp}\pm{\vec k}_{2\,\perp}|^2=\frac{\pi^2}{\lambda_p^2}
 \Big[\sin^2\theta_1+\sin^2\theta_2\\
 \mp 2\sin\theta_1\sin\theta_2\cos(\alpha_1-\alpha_2)\Big].
 \label{sum and diff}
\end{gather}
Let us assume that all deviation angles and differences of angles are small:
\begin{equation}
 \label{conditions}
 |\theta_{1,2}-\theta_0|\ll\theta_0\ll 1,\quad |\alpha_1-\alpha_2|\ll 1.
\end{equation}
In these approximations Eq. (\ref{sum and diff}) takes even much simpler form
\begin{equation}
 |{\vec k}_{1\,\perp}\pm{\vec k}_{2\,\perp}|^2\approx\frac{\pi^2}{\lambda_p^2}
 \Big[(\theta_1\mp\theta_2)^2\pm \theta_0^2(\alpha_1-\alpha_2)^2\Big].
 \label{sum and diff-approx}
\end{equation}

\subsection{Refractive index of the pump wave}

In a crystal, absolute values of the wave vectors $k_p$ and $k_{1,2}$ are determined by the corresponding refractive indices:
$k_1=k_2=\frac{\pi}{\lambda_p}n_o(2\lambda_p)$ and $k_p=\frac{2\pi}{\lambda_p}n_p(\lambda_p,\varphi_0,\alpha_p,\varphi_p )$, where  $n_o(\lambda)$ is the isotropic refractive index of the ordinary wave depending only on the light wavelength, and the pump refractive index is given by
\begin{gather}
 \nonumber
 n_p(\lambda_p,\varphi_p,\alpha_p,\varphi_0)=n_o(\lambda_p)n_e(\lambda_p) \Big\{ n_o^2(\lambda_p)\Big[\sin^2\varphi_p\sin^2\alpha_p\\
 \nonumber
 +
 (\sin\varphi_p\cos\varphi_0\cos\alpha_p+\cos\varphi_p\sin\varphi_0)^2\Big]+\\
 \label{ne-general}
 n_e^2(\lambda_p)(\cos\varphi_p\cos\varphi_0-\sin\varphi_p\sin\varphi_0\cos\alpha_p)^2\Big\}^{-1/2}.
\end{gather}
In this formula $n_e(\lambda_p)$ is the extraordinary-wave refractive index for the propagation direction along the minor axis of the polarization ellipse of a crystal.

As all deviations from the $z$-axis are assumed to be small, the refractive index of Eq. (\ref{ne-general}) can be expanded in powers of $\varphi_p$. With only two first terms of this expansion retained, the first part of the expression on the right-hand side of Eq. (\ref{Delta}) takes the form
\begin{gather}
\nonumber
 k_p-k_1-k_2=\frac{2\pi}{\lambda_p}\left[n_p(\lambda_p,\varphi_p,\alpha_p,\varphi_0)-n_o(2\lambda_p)\right]\approx\\
 \label{p-1-2}
 \frac{2\pi}{\lambda_p}[n_p(\varphi_0)-n_o +n_p^\prime(\alpha_p,\varphi_0)\varphi_p],
\end{gather}
where
\begin{equation}
  n_p(\varphi_0)\equiv n_p=n_p(\lambda_p,0,0,\varphi_0),\;n_o=n_o(2\lambda_p)
 \label{np(phi0)}
\end{equation}
and
\begin{gather}
 \nonumber
 n_p^\prime(\lambda_p,\alpha_p,\varphi_0)\equiv n_p^\prime=\left.\frac{\partial n_p(\lambda_p,\varphi_p,\alpha_p,\varphi_0)}{\partial\varphi_p}\right|_{\varphi_p=0}=\\
 \label{deriv}
 -\zeta(\lambda_p,\varphi_0)\cos\alpha_p.
\end{gather}
The term with the derivative of the refractive index $n_p^\prime$ in Eq. (\ref{p-1-2}) determines the well known spacial walk-off effects. Very often this term is omitted from consideration at all. Let us refer this simplification as corresponding to the ``No Walk-Off$"$ or NWO approximation. As shown below, for azimuthal entanglement, in the case of sufficiently well pronounced noncollinearity, the NWO approximation is reasonably good. But in a general case the walk-off term can be important because it takes into account anisotropy of birefringent crystals and breaks the axial symmetry of the SPDC process \cite{2007,2008}. A structure of the walk-off term is  discussed in the subsection 2.5. But before this, let us discuss the structure of other terms in the phase mismatch of Eq. (\ref{Delta}).

\subsection{The NWO parts of the phase mismatch, linear approximation}

\begin{figure}[h]
\centering\includegraphics[width=6cm]{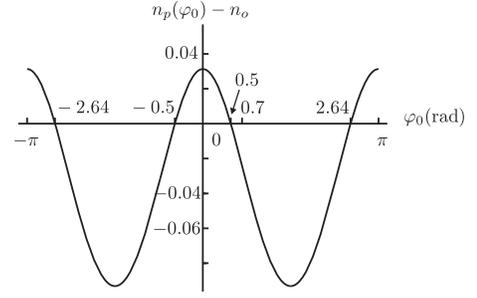}
\caption{{\protect\footnotesize {The difference of the refractive indices $n_p-n_o$ as a function of $\varphi_0$ for BBO crystal and the wavelength $\lambda_p=0.4047\,{\rm mkm}$.}}}\label{Fig3}
\end{figure}

The term $n_p(\varphi_0)-n_o$ in Eq. (\ref{p-1-2}) determines the dependence of the phase mismatch on the angle $\varphi_0$ between the optical axis and the pump propagation direction $0z$. For BBO crystal and the pump wavelength $\lambda_p=0.4047$ mkm (as, e.g., in the Ti:sapphire laser used in the work \cite{Kriv}) this dependence is shown in Fig. \ref{Fig3}.
The angle $\varphi_0^{\rm coll}=0.5$ corresponds to the collinear SPDC regime, and the non-collinear regime occurs at $0.5<|\varphi_0|<2.64$. Further numerical examples will be obtained at $\varphi_0=0.7$.

The second NWO part in the phase mismatch, $\frac{1}{2}\left({\vec k}_{1\,\perp}+{\vec k}_{2\,\perp}\right)^2$, is determined by Eq. (\ref{sum and diff}), and together with the expression of Eq. (\ref{p-1-2}) this gives
\begin{gather}
 \nonumber
\Delta_{\rm NWO}=\frac{\pi}{4n_o\lambda_p}
 \Big[(\theta_1+\theta_2)^2+8n_o(n_p-n_o)-\\
 \nonumber
  \theta_0^2(\alpha_1-\alpha_2)^2 \Big]\equiv
  \frac{\pi}{4n_o\lambda_p}\Big[(\theta_1+\theta_2)^2-4\theta_0^2-\theta_0^2(\alpha_1-\alpha_2)^2 \Big]\\
  \label{Delta-appr}
 \approx\frac{\pi}{n_o\lambda_p}\theta_0(\theta_1+\theta_2-2\theta_0),
\end{gather}
where $\theta_0$ is defined at last as
\begin{equation}
 \label{theta0}
 \theta_0=\sqrt{2n_o(n_o-n_p)}.
\end{equation}
Numerically, for BBO crystal and $\lambda_p=0.4047$ mkm, at $\varphi_0=0.7$, Eq. (\ref{theta0}) yields $\theta_0=0.28\approx 16^\circ$.

Note that, in fact, $\theta_0$ (\ref{theta0}) is defined in the transition from the first line in Eq. (\ref{Delta-appr}) to the second one, whereas the third line of Eq. (\ref{Delta-appr}) is the approximation based on the assumptions of Eq. (\ref{conditions}). In this approximation the dependence of $\Delta$ on the polar angles becomes linear, whereas the dependence on the difference of azimuthal angles remains quadratic. For this reason the term $\theta_0^2(\alpha_1-\alpha_2)^2$ is dropped in the last line of Eq. (\ref{Delta-appr}) as having a higher order in small difference of angles compared to the retained linear term $\theta_0(\theta_1+\theta_2-2\theta_0)$. Smallness of the dropped quadratic term is verified by its direct estimate in subsection 3.1 below.

Note also that, in accordance with Eq. (\ref{Delta-appr}), the constant part of the phase mismatch, independent of any variables, is $\Delta_0=-\pi\theta_0^2/4n_o\lambda_p$, and this gives the following expression for the constant term in the argument of the sinc-function in Eq. (\ref{Psi}):
\begin{equation}
 \label{phi}
 \phi=\frac{L\Delta_0}{2}=-\frac{\pi\theta_0^2L}{8n_0\lambda_p}.
\end{equation}
Estimated at the same values of parameters as indicated above and $L=0.5$ cm, Eq. (\ref{phi}) gives $\phi\approx -900$, which exceeds significantly the values $\phi= -2.3$ or $\phi= -4$ of Refs. \cite{Giovannini} and  \cite{Miatto}. At the same other parameters as used above these small values of $\phi$ correspond to very small degree of noncollinearity, $\theta_0\leq 0.17\approx 1^\circ$. As will be shown below, this large difference with the case under consideration ( $\theta_0=0.28\approx 16^\circ$)  explains a large differences in the predicted degree of azimuthal entanglement.

\subsection{Evaluation of the ``walk-off$"$ term}
\begin{figure}[h]
\centering\includegraphics[width=6cm]{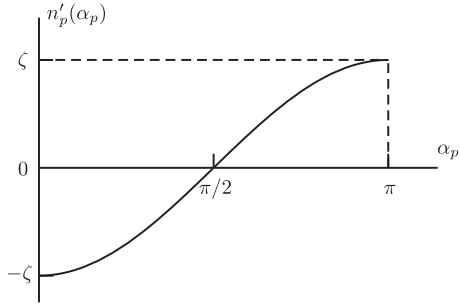}
\caption{{\protect\footnotesize {Derivative of the refractive index $n_p^\prime$ (\ref{deriv}) as a function of the pump azimuthal angle $\alpha_p$.}}}\label{Fig4}
\end{figure}

Let us consider now the ``walk-off$"$ term in the phase mismatch determined by the first-order derivative of the refractive index in Eq. (\ref{p-1-2}),
\begin{equation}
 \label{walk-off-term}
 \frac{2\pi}{\lambda_p}n_p^\prime\varphi_p=-\frac{2\pi}{\lambda_p}\zeta(\lambda_p,\varphi_0)\varphi_p\cos\alpha_p.
\end{equation}
In the case of a BBO crystal and the same parameters as indicated above, $\zeta(\lambda_p=0.4047,\varphi_0=0.7)\approx 0.12$.  The dependence $n_p^\prime\propto\cos\alpha_p$ characterizes anisotropy of the walk-off term. As shown in Fig. \ref{Fig4}, anisotropy is maximally pronounced at $\alpha=0$ or $\pi$ (when ${\vec k}_p\in (xz)$) and turns zero at $\alpha=\pi/2$ (when ${\vec k}_p\in(yz)\perp (xz)$), where $(xz)$ is the plane containing the crystal optical axis (see also the definitions of $\alpha$ and $\varphi_0$ in Fig. \ref{Fig1}).

For further analysis both $\varphi_p$ and $\cos\alpha_p$ have to be expressed in terms of the photon angles $\theta_{1,2}$ and $\alpha_{1,2}$ in a free space after the crystal, which can be done with the help of Eq. (\ref{transv-conserv}) (transverse-momentum conservation rule).

At first, note that the tangent components of wave vectors are continuous at the crystal-vacuum boundary. For this reason the left- and right-hand sides of Eq. (\ref{transv-conserv}) can be evaluated, correspondingly, inside and outside of the crystal to give for the squared terms on both sides of this equation
\begin{equation}
  \label{varphi-p}
  k_p^2\sin^2\varphi_p\approx k_p^2\varphi_p^2=({\vec k}_{1\,\perp}+{\vec k}_{2\,\perp})^2
\end{equation}
or
\begin{equation}
  \label{varphi-p-2}
  \varphi_p=\frac{\lambda_p}{2\pi n_p}|{\vec k}_{1\,\perp}+{\vec k}_{2\,\perp}|,
\end{equation}
with known  expressions for  $({\vec k}_{1\,\perp}+{\vec k}_{2\,\perp})^2$ and, hence, $|{\vec k}_{1\,\perp}+{\vec k}_{2\,\perp}|$ in terms of $\theta_{1,2}$ and $\alpha_{1,2}$ (\ref{sum and diff}), (\ref{sum and diff-approx}).

Another way of using Eq. (\ref{transv-conserv}) consists in  projecting its both sides on the direction perpendicular the vector ${\vec k}_{p\,\perp}$. Then, clearly, the left-hand side of Eq. (\ref{transv-conserv}) gives zero, and the right-hand side gives the equation for finding $\cos\alpha_p$:
\begin{equation}
 \label{eq foe cos alpha-p}
 \sin\theta_1\sin(\alpha_1-\alpha_p)-\sin\theta_2\sin(\alpha_2-\alpha_p)=0,
\end{equation}
Solution of this equation is given by
\begin{gather}
 \nonumber
 \cos\alpha_p=\\
 \label{cos alp-p}
 \frac{\sin\theta_1\cos\alpha_1-\sin\theta_2\cos\alpha_2}{[\sin^2\theta_1+\sin^2\theta_2-2\sin\theta_1\sin\theta_2\cos(\alpha_1-\alpha_2)]^{1/2}}.
\end{gather}
The denominator of the expression is seen to coincide with the angular part of $|{\vec k}_{1\,\perp}-{\vec k}_{2\,\perp}|$ of Eq. (\ref{sum and diff}). As for the numerator of the fraction in Eq. (\ref{cos alp-p}), as previously done, it can be expanded in powers $|\theta_1-\theta_2|$ and $|\alpha_1-\alpha_2|$ with only the lowest-order  (linear) terms to be retained. Then the final expression for $\cos\alpha_p$ takes the form
\begin{gather}
 \nonumber
 \cos\alpha_p=\frac{\pi/\lambda_p}{|{\vec k}_{1\,\perp}+{\vec k}_{2\,\perp}|}\times\\
 \label{cos alp-p-final}
 \big[(\theta_1-\theta_2)\cos\alpha_0-\theta_0\sin\alpha_0(\alpha_1-\alpha_2)\big],
\end{gather}
where
\begin{equation}
 \label{alpha-0}
 \alpha_0=\frac{\alpha_1+\alpha_2}{2}.
\end{equation}
In contrast to the difference of azimuthal angle $\alpha_1-\alpha_2$, which is assumed to be small, their sum and a half-sum $\alpha_0$ can change in rather wide ranges, $\frac{\pi}{2}\geq\alpha_0\geq -\frac{\pi}{2}$. With obtained expressions (\ref{varphi-p-2}) and (\ref{cos alp-p-final}), the contribution (\ref{walk-off-term}) of the walk-off term into the phase mismatch $\Delta$ takes the form
\begin{equation}
 \label{varp x cos}
 -\frac{\pi\zeta}{\lambda_p n_p}\big[(\theta_1-\theta_2)\cos\alpha_0-\theta_0\sin\alpha_0(\alpha_1-\alpha_2)\big].
\end{equation}
Note that this expression can seem to be antisymmetric with respect to the variable transposition $\theta_1\rightleftharpoons\theta_2$, $\alpha_1\rightleftharpoons\alpha_2$, which cannot be true. In fact, however, it should be kept in mind that, in accordance with the definition of the angles $\alpha_{1,2}$ shown in Fig. \ref{Fig2}, the variable transposition must be accompanied by the shift of both angles $\alpha_1$ and $\alpha_2$ for $\pi$, which changes the signs of $\cos\alpha_0$ and $\sin\alpha_0$ and provides the symmetry of the expression (\ref{varp x cos}).

\subsection{Final expressions for the wave function and its double-Gaussian representation}
Summation of all derived results (\ref{Psi}),(\ref{Delta-appr}), (\ref{sum and diff-approx}), (\ref{walk-off-term}), (\ref{varp x cos}) gives the following general expression for the biphoton angular wave function
\begin{gather}
  \nonumber
  \Psi\propto\exp\left\{-\frac{(\theta_1-\theta_2)^2+ \theta_0^2(\alpha_1-\alpha_2)^2}{2\Delta\theta_p^2}\right\}\times\\
  \nonumber
  {\rm sinc}\left\{\frac{1}{2\Delta\theta_L}\Big[\theta_0(\theta_1+\theta_2-2\theta_0)-\right.\\
  \left.\frac{n_o}{n_p}\zeta\Big(\cos\alpha_0(\theta_1-\theta_2)-\sin\alpha_0\theta_0(\alpha_1-\alpha_2)\Big)\Big]\right\},
 \label{Psi-fin}
\end{gather}
where $\Delta\theta_p=\lambda_p/\pi w$ and  $\Delta\theta_L=n_o\lambda_p/\pi L$.

As mentioned above, in principle, the walk-off terms make angular biphoton wave function axially asymmetric, which is seen in its dependence not only on the difference of azimuthal angles $\alpha_1-\alpha_2$ but also on their sum via $\alpha_0$ (\ref{alpha-0}). The symmetry returns in the NWO approximation when the expression (\ref{Psi-fin}) takes the form
\begin{gather}
 \nonumber
 \Psi_{\rm NWO}\propto\exp\left\{-\frac{(\theta_1-\theta_2)^2+ \theta_0^2(\alpha_1-\alpha_2)^2}{2\Delta\theta_p^2}\right\}\\
  \label{Psi-NWO}
  \times{\rm sinc}\left\{\frac{1}{2\Delta\theta_L}\Big[\theta_0(\theta_1+\theta_2-2\theta_0)\Big]\right\}.
\end{gather}
In addition to the dependence only on the difference of azimuthal angles $\alpha_1-\alpha_2$ the remarkable feature of the NWO approximation consists in factorization of the dependencies on polar and azimuthal angles. In other words, the NWO wave function (\ref{Psi-NWO}) takes the form of a product of two factors $\Psi_{\rm NWO}=\Psi_{\rm NWO}^{\rm pol}\times\Psi_{\rm NWO}^{\rm az}$ with $\Psi_{\rm NWO}^{\rm pol}$ and $\Psi_{\rm NWO}^{\rm az}$ depending,  correspondingly, only on polar and only on azimuthal angles
\begin{gather}
 \nonumber
 \Psi_{\rm NWO}^{\rm pol}(\theta_1,\theta_2)\propto\exp\left\{-\frac{(\theta_1-\theta_2)^2}{2\Delta\theta_p^2}\right\}\\
 \label{Psi-pol-NWO}
 \times{\rm sinc}\left\{\frac{1}{2\Delta\theta_L}\Big[\theta_0(\theta_1+\theta_2-2\theta_0)\Big]\right\}
\end{gather}
 and
\begin{equation}
 \label{Psi-az-NWO}
 \Psi_{\rm NWO}^{\rm az}(\alpha_1,\alpha_2)\propto\exp\left\{-\frac{
 \theta_0^2(\alpha_1-\alpha_2)^2}{2\Delta\theta_p^2}\right\}.
\end{equation}
In contrast, beyond the NWO approximation, with the walk-off terms taken into account, there is no factorization in the wave function of Eq. (\ref{Psi-fin}) for parts depending on polar and azimuthal variables, and no axial symmetry.

\begin{figure}[h]
\centering\includegraphics[width=6cm]{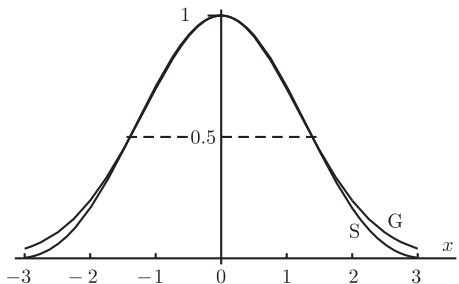}
\caption{{\protect\footnotesize {The functions sinc$^2(x)$ (S) and $\exp(-0.359 x^2)$ (G).}}}\label{Fig5}
\end{figure}

The next step of simplifications in the general expression (\ref{Psi-fin}) consists in the replacement of the squared sinc-function by the appropriately chosen Gaussian function:
\begin{equation}
 \label{sinc-gauss}
 {\rm sinc}^2(x)\rightarrow\exp(-0.359x^2),
\end{equation}
where 0.359 is the best fitting parameter, and quality of the replacement is illustrated by the picture of Fig. \ref{Fig5}.
This modeling gives the following double-Gaussian representation for the polar-azimuthal angular distribution of the biphoton probability density
\begin{gather}
 \nonumber
 |\Psi(\theta_1,\alpha_1;\theta_2,\alpha_2)|^2\propto\exp\left[  -\frac{(\theta_1-\theta_2)^2+\theta_0^2(\alpha_1-\alpha_2)^2}{\Delta\theta_p^2}\right]\times\\
 \nonumber
 \exp\Bigg\{-\frac{0.395}{4\Delta\theta_L^2}\Big[\theta_0(\theta_1+\theta_2-2\theta_0)-\\
 \label{double-Gauss}
 \frac{n_o}{n_p}\zeta\Big(\cos\alpha_0(\theta_1-\theta_2)-\sin\alpha_0\theta_0(\alpha_1-\alpha_2)\Big)\Big]^2\Bigg\}.
\end{gather}
Note that the arguments of both exponents are quadratic in variables and, thus, the linear approximation in the argument of the sinc-function in Eq. (\ref{Psi-fin}) fits perfectly the quadratic dependencies in the Gaussian functions (\ref{double-Gauss}).

\section{Entanglement}

As mentioned in the Introduction, the main goal of the present consideration consists in characterization and evaluation of the purely azimuthal entanglement of biphotons independently of entanglement in the second degree freedom, i.e., in polar angles.

\subsection{Coincidence and single-particle widths}

As known \cite{2004, 2006}, the degree of entanglement of bipartite states with continuous variables can be found by means of single-particle and coincidence measurements. Such measurements can be used to plot curves of the corresponding distributions of numbers of registered particles in dependence on the corresponding variables. The ratio of widths of the single-particle and coincidence distributions is the parameter $R$ characterizing the degree of entanglement. Mathematically, this parameter is defined as the ratio of the widths of the curves of unconditional and conditional probability densities. For double-Gaussian wave functions this parameter is known to coincide exactly with the Schmidt parameter $K=1/Tr(\rho_r^2)$ where $\rho_r$ is the reduced density matrix of the bipartite state. It's clear also that if particles in a bipartite state have two degrees of freedom, then for finding probability distributions and entanglement in one of these degrees of freedom, one has to integrate the total two-degrees-of-freedom distribution over variables characterizing the second degree of freedom.

\begin{figure}[h]
\centering\includegraphics[width=6cm]{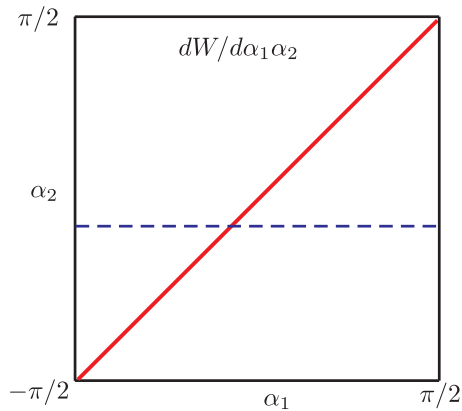}
\caption{{\protect\footnotesize {Density plot of the biphoton azimuthal distribution  (\ref{azim-distr})}}}\label{Fig6}
\end{figure}

Complete independence of azimuthal and polar-angle parts in the biphoton angular wave function occurs in the NWO approximation (\ref{Psi-NWO}), when the distribution of the biphoton azimuthal probability density is given by
\begin{equation}
 \label{azim-distr}
 \frac{dW}{d\alpha_1d\alpha_2}\propto\exp\left[-\frac{\theta_0^2(\alpha_1-\alpha_2)^2}{\Delta\theta_p^2}\right]
\end{equation}
with the only additional restriction
\begin{equation}
 \label{restrict}
 \frac{\pi}{2}\geq\alpha_0\equiv\frac{\alpha_1+\alpha_2}{2}\geq-\frac{\pi}{2}.
\end{equation}
The density plot of this distribution is shown schematically in the picture of Fig. \ref{Fig6}.
In the map ($\alpha_1,\alpha_2$), this is a very narrow ridge of a unit hight, going along the diagonal $\alpha_1=\alpha_2$ from $\alpha_1=\alpha_2=-\pi/2$ and up to $\alpha_1=\alpha_2=\pi/2$. The coincidence distribution in $\alpha_1$ has to be measured with one detector counting photons only at some given value of $\alpha_2$ and the second detector scanning, in the map ($\alpha_1,\alpha_2$) - along the horizontal  line $\alpha_2=const$ (the blue dashed line in Fig. \ref{Fig6}), and with only joint signals registered. In single-particle measurements one has to use only one detector scanning horizontally (in the map ($\alpha_1,\alpha_2$)) and registering all photons, independently of and at all possible values of the second-photon azimuthal angle $\alpha_2$. Clearly, the coincidence and single-particle widths in such schemes of measurements are equal to
\begin{equation}
 \label{coi-single}
 \Delta\alpha_1^{(c)}=\frac{\Delta\theta_p}{\theta_0},\quad \Delta\alpha_1^{(s)}=\pi,
\end{equation}
and they correspond to the entanglement parameter
\begin{equation}
 \label{R}
 R=\frac{\Delta\alpha_1^{(s)}}{\Delta\alpha_1^{(c)}}=\frac{\pi\theta_0}{\Delta\theta_p}=\pi^2\theta_0\frac{ w}{\lambda_p}\sim 10^4.
\end{equation}

Beyond the NWO approximation polar and angular variables are not separated in the total wave function (\ref{Psi-fin}). In this case the bipartite azimuthal probability density can be obtained from the squared wave function of Eq. (\ref{double-Gauss}) by means of integration over the polar angles $\theta_1$ and $\theta_2$, or equivalently, at first over $\theta_1+\theta_2$ and then over $\theta_1-\theta_2$. An important feature of this integration is related to the structure of the second exponential function in Eq. (\ref{double-Gauss}). Though it depends on both $\theta_1+\theta_2$ and $\theta_1-\theta_2$, its integration over $\theta_1+\theta_2$ from $-\infty$ to $+\infty$ gives just a number independent of any variables. This is a direct consequence of the linear approximation used for evaluation of the argument of the sinc-function in Eq. (\ref{Psi-fin}). The second integration, over $\theta_1-\theta_2$, is carried out equally easy to give the expression for the azimuthal biphoton distribution coinciding with that derived in the NWO approximation (\ref{azim-distr}) and shown in Fig. \ref{Fig6}. Hence, all estimates of the azimuthal coincidence and single-particle widths (\ref{coi-single}), and of the parameter of azimuthal entanglement $R$ (\ref{R}) remain valid even beyond the NWO approximation when the walk-off term is taken into account in the phase mismatch in the linear approximation in $\theta_1\pm\theta_2$ and $\alpha_1-\alpha_2$. Note that this proof is valid explicitly only for the described derivation of the parameter $R$, because the widths in its definition (\ref{R}) are defined as widths of the probability distributions $dW/d\alpha_1d\alpha_2|_{\alpha_2=const}$ and $\int d\alpha_2dW/d\alpha_1d\alpha_2$, rather than widths of wave functions. In contrast to this the described in the following two subsection alternative derivations deal with the wave biphoton azimuthal function and are valid only in the NWO approximation.

The azimuthal coincidence width of Eq. (\ref{coi-single}) can be used to estimate explicitly (as promised above) the degree of smallness of the term in the phase mismatch (\ref{Delta-appr}) quadratic  in the difference of azimuthal angles and dropped in the linear-approximation. Contribution of this term into the argument of the sinc-function in Eq. (\ref{Psi-fin}) would be given by
\begin{equation}
 \label{diffr-L}
\frac{\theta_0^2(\alpha_1-\alpha_2)^2}{8\Delta\theta_L}\sim \frac{\Delta\theta_p^2}{8\Delta\theta_L}=\frac{L}{8n_oL_D}\ll 1,
\end{equation}
where $L_D=\pi w_p^2/\lambda_p$ is the diffraction length (or Rayleigh range) of the pump which is assumed always to be much longer than the crystal length $L$.

\subsection{Schmidt-mode analysis}

The same results as described above can be obtained in the Schmidt-decomposition formalism. Azimuthal Schmidt modes can be found with the help of a slight modification in the azimuthal wave function $\Psi_{\rm NWO}$ (\ref{Psi-NWO}). Let the restriction $|\alpha_0|\leq\pi/2$ (\ref{restrict}) be imitated (replaced) by an additional Gaussian factor
$\exp\left[-(\alpha_1+\alpha_2)^2/8\pi^2\right]$,  which reduces the azimuthal wave function to the standard double-Gaussian form of Ref. \cite{2006}
\begin{equation}
 \label{double-Gauss-gen}
 {\widetilde\Psi}=N
 \exp\left(-\frac{(\alpha_1+\alpha_2)^2}{2a^2}\right)\exp\left(-\frac{(\alpha_1-\alpha_2)^2}{2b^2}\right)
\end{equation}
with $N=\sqrt{2/\pi ab}$, $a=2\pi$, $b=\Delta\theta_p/\theta_0$ and, evidently, $a\gg b$ . For such wave functions their Schmidt decomposition and Schmidt modes are known \cite{Uren,2009,CP}:
\begin{equation}
 \label{decomp}
 {\widetilde\Psi}(\alpha_1,\alpha_2)=\sum_n\sqrt{\lambda_n}\psi_n(\alpha_1)\psi_n(\alpha_2),
\end{equation}
where
\begin{equation}
 \label{lambda-n}
 \lambda_n=\frac{4ab}{(a+b)^2}\left(\frac{a-b}{a+b}\right)^{2n},
\end{equation}
$\psi_n$ are Schmidt modes
\begin{equation}
 \label{modes}
 \psi_n=\left(\frac{2}{ab}\right)^{1/4}u_n\left(\frac{\sqrt{2}\,\alpha}{\sqrt{ab}}\right),
\end{equation}
and $u_n(x)$ are the Hermite-Gaussian functions
\begin{equation}
 \label{H-G}
 u_n(x)=(2^nn!\sqrt{\pi})^{-1/2}e^{-x^2/2}H_n(x).
\end{equation}
The constants $\lambda_n$ (\ref{lambda-n}) determine the Schmidt parameter $K$ characterizing both the degree of entanglement and the effective dimensionality
of the Hilbert space
\begin{equation}
 \label{K}
 K=\frac{1}{\sum_n\lambda_n^2}=\frac{a^2+b^2}{2ab}\approx \frac{a}{2b}=\frac{\pi\theta_0}{\Delta\theta_p}=\frac{\pi^2\theta_0\,w}{\lambda_p}.
\end{equation}
Comparison with Eq. (\ref{R}) shows that the entanglement parameters $K$ and $R$ are identically equal to each other and, thus, the derivation in terms of Schmidt modes confirms perfectly the found above very high level of azimuthal entanglement .

\subsection{${\rm OAM}$ analysis}

Let us consider only the azimuthal part (\ref{Psi-az-NWO}) of the biphoton angular wave function in the NWO approximation. The function $\exp\left[-(\alpha_1-\alpha_2)^2/2[\Delta\alpha^{(c)}]^2\right]$ can be expanded in a series of products of adjoint OAM eigenfunctions $e^{il\alpha_1}$ and $e^{-il\alpha_2}$, where $l=0,\,\pm 1,\,\pm 2, ...$ are the OAM eigenvalues.
\begin{equation}
 \label{OAM-decomp}
 \Psi_{\rm NWO}^{\rm az}\propto\sum_l\,C_l\,e^{il\alpha_1}\times e^{-l\alpha_2}
\end{equation}
The expansion coefficients $C_l$ can be easily found to be proportional to $\exp\left[-l^2[\Delta\alpha^{(c)}]^2/2\right]$ with $\Delta\alpha^{(c)}$ being the coincidence width of the first Eq. (\ref{coi-single}). As these coefficients are even in $l$, only the real part of the product of OAM eigenfunctions gives nonzero contribution into the sum over $l$, i.e.,
\begin{gather}
\nonumber
e^{il\alpha_1}\times e^{-il\alpha_2}\equiv\cos[l(\alpha_1-\alpha_2)]+i\sin[l(\alpha_1-\alpha_2)]\\
\nonumber
\Rightarrow\cos[l(\alpha_1-\alpha_2)]=\cos l\alpha_1\cos l\alpha_2+\sin l\alpha_1\sin l\alpha_2.
\end{gather}
As a result, the normalized azimuthal wave function takes the form of the OAM Schmidt decomposition
\begin{gather}
 \nonumber
 \Psi(\alpha_1,\alpha_2)=\sum_l\sqrt{\lambda_l^{(\rm OAM)}}\,\Big[\psi_{l\,{\rm OAM}}^{(\rm cos)}(\alpha_1)\psi_{l\,{\rm OAM}}^{(\rm  cos)}(\alpha_2)\\
 \label{Schm-dec-OAM}
 +\psi_{l\,{\rm OAM}}^{(\rm sin)}(\alpha_1)\psi_{l\,{\rm OAM}}^{(\rm sin)}(\alpha_2)\Big],
\end{gather}
where
\begin{equation}
 \label{Schm-eigen-OAM}
 \lambda_l^{(\rm OAM)}=\frac{\Delta\alpha^{(c)}}{2\sqrt{\pi}}\exp\left(-l^2[\alpha^{(c)}]^2\right),
 \end{equation}
and the OAM Schmidt modes are given by
\begin{equation}
 \label{Schm-mod-OAM}
 \psi_{l\,{\rm OAM}}^{(\rm cos)}(\alpha)=\sqrt{\frac{2}{\pi}}\cos l\alpha,\;
 \psi_{l\,{\rm OAM}}^{(\rm sin)}(\alpha)=\sqrt{\frac{2}{\pi}}\sin l\alpha
\end{equation}
with $|\alpha|\leq\pi/2$.

All OAM Schmidt modes are twice degenerate, owing to which the normalization has the form $2\sum_l\ \lambda_l^{(\rm OAM)}=1$, which is checked by means of summation substituted by integration over $l$. The OAM Schmidt entanglement parameter is defined as the inverse double sum of squared $\lambda_l^{(\rm OAM)}$:
\begin{gather}
 \label{K-OAM}
 K_{az}^{(\rm OAM)}=\frac{1}{2\sum_l\left(\lambda_l^{(\rm OAM)}\right)^2}=
  \frac{2\sqrt{2\pi}\,\theta_0 w}{\lambda_p},
\end{gather}
in a sufficiently good agreement with the above-found parameters $K$ (\ref{K}) and $R$ (\ref{R}).

\subsection{Discussion}
Thus, found degree of azimuthal entanglement is rather high and, besides, its features are somewhat unusual. In particular, as seen from Eqs. (\ref{R}), (\ref{K}) and (\ref{K-OAM}), the degree of azimuthal entanglement is determined only by the pump angular width $\Delta\theta_p$ and does not depend of the the crystal length $L$. Of course, this result can be valid only in a restricted variation range of $L$. One restriction is given by Eq. (\ref{diffr-L}). It arises from the condition that the term $\theta_0^2(\alpha_1-\alpha_2)^2/8\Delta\theta_L$ in the argument of the sinc-function is small and can be dropped in the linear approximation. Another restriction arises from the linearization condition of the difference $(\theta_1+\theta_2)^2-4\theta_0^2$. Approximation of this expression by $4\theta_0(\theta_1+\theta_2-2\theta_0)$ assumes that the term $(\theta_1+\theta_2-2\theta_0)^2$ gives only a negligibly small contribution to the argument. As follows from Eqs. (\ref{Psi-NWO}), (\ref{Psi-pol-NWO}) $|\theta_1+\theta_2-2\theta_0|\sim\Delta\theta_L/\theta_0$, which gives the following estimate of the term which has been dropped
\begin{equation}
 \label{lin-appr-validity}
 \frac{(\theta_1+\theta_2-2\theta_0)^2}{\Delta\theta_L}\sim\frac{\Delta\theta_L}{\theta_0^2}
=\frac{n_o\lambda_p}{\pi L\theta_0^2}\ll 1.
\end{equation}
In terms of the crystal length $L$ the condition (\ref{lin-appr-validity}) gives
\begin{equation}
 \label{restr for L}
 L\gg\frac{n_o\lambda_p}{\pi\theta_0^2}\approx 2.78\,{\rm mkm},
\end{equation}
where the last estimate is obtained for all the same parameters as used throughout the paper, including $\theta_0=0.28$. As seen from Eq. (\ref{restr for L}) the found condition is fulfilled practically for any realizable values of the crystal length $L$. But at smaller degrees of noncollinearity (at smaller values of $\theta_0$) the restriction (\ref{restr for L}) becomes somewhat more serious. And, definitely, the restrictions (\ref{lin-appr-validity}) excludes the case $\theta_0=0$, which corresponds to collinear SPDC regime when linearization of the phase mismatch is impossible at all.

\section{Multichannel Schimdt-type decomposition}

\begin{figure}[h]
\centering\includegraphics[width=8.5cm]{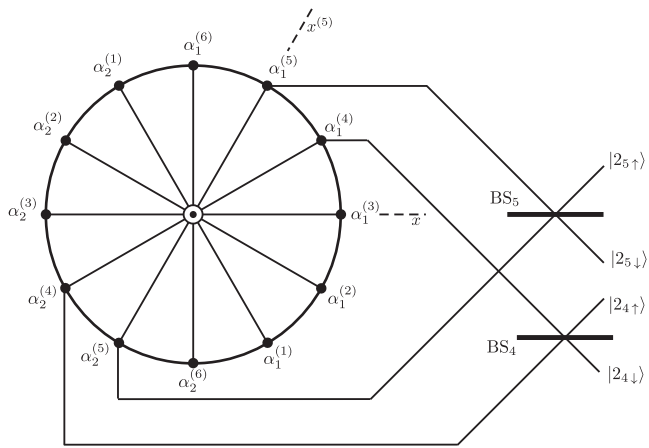}
\caption{{\protect\footnotesize {A scheme of a possible experiment for getting the multichannel Schmidt-type decomposition of noncollinear biphoton states revealing partially their high resource of azimuthal entanglement.}}}\label{Fig7}
\end{figure}

The found above very high values of the parameters $R$ and $K$ indicate a very high level of the entanglement resource accumulated in the azimuthal variables of noncollinear biphotons and, on the other hand, they indicate serious problems with any attempts of physical separation of azimuthal Schmidt modes. Indeed, the amount of more or less equally important terms in the Schmidt decomposition (\ref{decomp}) is on the  order of $K\gg 1$. High-order Hermit-Gaussian functions (\ref{H-G}) are very rapidly oscillating, and neighboring Schmidt modes are very similar to each other. For these reasons, the task of separation of such modes in experiments looks hopeless. This raises a question: are there any ways of using a very high resource of azimuthal entanglement, though at least partially. One scheme for construction the Schmidt-type separation of orthogonal modes is shown in Fig. \ref{Fig7}. The main idea consists in collecting pairs of photon from a series of different planes $(x_n,z)$ with axes $0x_n\in (xy)$ directed at angles $\alpha^{(n)}$ with respect to the horizontal $x$-axis.
In each plane $(x_n,z)$ photons arise at opposite ends of the corresponding diameter around azimuthal angles $\alpha_1^{(n)}=\alpha^{(n)}$ and $\alpha_2^{(n)}=\alpha^{(n)}+\pi$. Black spots at the ends of diameters can be considered as symbolizing receiving fibers, sizes of which are assumed to be larger than the ring thickness ($\sim\Delta\theta_L/\theta_0$) to collect all photons with different polar angles $\theta$ at any given $\alpha_n$. On the other hand, the azimuthal distances between neighboring planes $\alpha^{(n+1)}-\alpha^{(n)}$ must be mach larger than sizes of receiving fibers and than the coincidence azimuthal width $\Delta\theta_p/\theta_0$. This last condition provides orthogonality of states arising in different planes $(x_n,z)$. As a whole, the state vector of all manifold of photons to be collected in the scheme of Fig. \ref{Fig7} is given by
\begin{equation}
 \label{N-st-vect}
 \ket{\Psi}=\frac{1}{\sqrt{N}}\sum_{n=1}^N\ket{1_{\alpha_1^{(n)}},1_{\alpha_2^{(n)}} }.
\end{equation}
After collection, each pair of photons from each given plane $(x_n,z)$ has to be sent to its own beam splitter which is the Hong-Ou-Mandel transformer \cite{HOM} directing united unsplit pairs of photon either to the up- or down-channels. The state vector of such transformed states takes the form
 \begin{equation}
 \label{N-sep-st-vect}
 \ket{\Psi}_{final}=\frac{1}{\sqrt{2N}}\sum_{n=1}^N\Big(\ket{2_{\uparrow\,n}}-\ket{2_{\downarrow\,n}}\Big),
\end{equation}
where the arrows $(\uparrow)$ and $(\downarrow)$ indicate the up- and down-channels and $n$ indicates the plane ($x_nz$) from which pairs of photons arrive to the $n^{th}$ beam splitter ${\rm BS_n}$. Two such pairs of channels are shown in Fig \ref{Fig7}. Altogether, the state  (\ref{N-sep-st-vect}) describes a physically separated multichannel Schmidt-type decomposition. In this state each pair of SPDC photons can appear with equal probability $\lambda_n=1/2N$ in one and only one (but arbitrary) of $2N$ channels. The degree of entanglement in such state  grows with growing amount of channels.  The Schmidt parameter $K$ and entropy of the reduced density matrix equal to
\begin{equation}
 \label{K-Sr-mult}
  K=2N,\quad S_r=-2\sum_{n=1}^N\lambda_n\log_2\lambda_n=1+\log_2N.
\end{equation}

\section{Conclusion}

Thus, entanglement of noncollinear biphoton states in azimuthal angles of photon wave vectors is considered, and the degree of azimuthal entanglement is found to be extremely high. The degree of azimuthal entanglement is evaluated by three methods: (1) by finding the parameter $R$ given by the ratio of the single-particle to coincidence widths of the angular distributions, (2) via the Schmidt parameter $K$ found for a model double-Gaussian wave function of two azimuthal angles and (3) in terms of the OAM analysis, in the frame of which the OAM Schmidt modes and decomposition are defined and found, as well as the OAM Schmidt number $K_{\rm az}^{\rm OAM}$ . All three methods are found to give the same estimate for the degree of entanglement, which is found to be determined by the ratio of the pump waist to its wavelength times the opening angle of the SPDC emission cone, $\theta_0\,w/\lambda_p$. For reasonably chosen values of all parameters the degree of entanglement and the effective dimensionality of the Hilbert space are found to be on the order of $10^4\gg 1$. A scheme is suggested for this very high resource of azimuthal entanglement to be seen experimentally, at least partially.

\section*{Acknowledgement}
The work is supported by the Russian Science Foundation, grant 14-02-01338


\bibliography{Text}

\end{document}